\documentclass[preprint]{vgtc}               

\graphicspath{{figures/}{pictures/}{images/}{./}} 
\usepackage{times}                     

\usepackage{tabu}                      
\usepackage{booktabs}                  
\usepackage{lipsum}                    
\usepackage{mwe}                       
\usepackage{listings}
\usepackage{balance}
\usepackage{color}
\definecolor{codegreen}{rgb}{0,0.6,0}
\definecolor{codegray}{rgb}{0.5,0.5,0.5}
\definecolor{codepurple}{rgb}{0.58,0,0.82}
\definecolor{backcolour}{rgb}{0,0,0}

\usepackage{mathptmx}                  
\usepackage{url}
\usepackage{enumitem}
\usepackage{tcolorbox}
\setlist{noitemsep,parsep=0pt,partopsep=0pt, leftmargin=10pt} 

\onlineid{7043}
\newcommand{\revise}[1]{{\color{black} #1}}

\vgtccategory{Research}

\vgtcinsertpkg

\newcommand{\eg}{{\it e.g.,\ }}

\NewDocumentCommand{\codeword}{v}{%
\texttt{\textcolor[HTML]{386594}{#1}}%
}

\newcommand{\tool}{Data Director~}
\newcommand{\toole}{Data Director}

\title{From Data to Story: Towards Automatic Animated Data Video Creation with LLM-based Multi-Agent Systems}

\usepackage{scalerel}
\usepackage{tikz}
\usetikzlibrary{svg.path}

\definecolor{orcidlogocol}{HTML}{A6CE39}
\tikzset{
  orcidlogo/.pic={
    \fill[orcidlogocol] svg{M256,128c0,70.7-57.3,128-128,128C57.3,256,0,198.7,0,128C0,57.3,57.3,0,128,0C198.7,0,256,57.3,256,128z};
    \fill[white] svg{M86.3,186.2H70.9V79.1h15.4v48.4V186.2z}
                 svg{M108.9,79.1h41.6c39.6,0,57,28.3,57,53.6c0,27.5-21.5,53.6-56.8,53.6h-41.8V79.1z M124.3,172.4h24.5c34.9,0,42.9-26.5,42.9-39.7c0-21.5-13.7-39.7-43.7-39.7h-23.7V172.4z}
                 svg{M88.7,56.8c0,5.5-4.5,10.1-10.1,10.1c-5.6,0-10.1-4.6-10.1-10.1c0-5.6,4.5-10.1,10.1-10.1C84.2,46.7,88.7,51.3,88.7,56.8z};
  }
}

\newcommand\orcidicon[1]{\href{https://orcid.org/#1}{\mbox{\scalerel*{
\begin{tikzpicture}[yscale=-1,transform shape]
\pic{orcidlogo};
\end{tikzpicture}
}{|}}}}

\author{
Leixian Shen~\orcidicon{0000-0003-1084-4912}
\thanks{e-mail: lshenaj@connect.ust.hk}\\ %
\scriptsize \parbox{1.35in}{The Hong Kong University of Science and Technology, Hong Kong SAR, China} %
\and Haotian Li~\orcidicon{0000-0001-9547-3449}
\thanks{e-mail: haotian.li@connect.ust.hk}\\ %
\scriptsize \parbox{1.35in}{The Hong Kong University of Science and Technology, Hong Kong SAR, China} %
\and Yun Wang~\orcidicon{0000-0003-0468-4043}\thanks{e-mail: wangyun@microsoft.com}\\ %
\scriptsize {Microsoft,}\\\scriptsize {Beijing, China} %
\and Huamin Qu~\orcidicon{0000-0002-3344-9694}\thanks{e-mail: huamin@cse.ust.hk}\\ %
\scriptsize \parbox{1.35in}{The Hong Kong University of Science and Technology, Hong Kong SAR, China} %
}

\abstract{
Creating data stories from raw data is challenging due to humans' limited attention spans and the need for specialized skills. 
Recent advancements in large language models (LLMs) offer great opportunities to develop systems with autonomous agents to streamline the data storytelling workflow.
Though multi-agent systems have benefits such as fully realizing LLM potentials with decomposed tasks for individual agents, designing such systems also faces challenges in task decomposition, performance optimization for sub-tasks, and workflow design. 
To better understand these issues, 
we develop \toole, an LLM-based multi-agent system designed to automate the creation of animated data videos, a representative genre of data stories.
\tool interprets raw data, breaks down tasks, designs agent roles to make informed decisions automatically, and seamlessly integrates diverse components of data videos. 
A case study demonstrates \toole's effectiveness in generating data videos. 
Throughout development, we have derived lessons learned from addressing challenges, guiding further advancements in autonomous agents for data storytelling. We also shed light on future directions for global optimization, human-in-the-loop design, and the application of advanced multimodal LLMs.

}

\keywords{Data Storytelling, LLM, Multi-Agent, Data Video}

\begin{document}


\firstsection{Introduction}

\maketitle

The rapid growth of data assets has driven advancements in various domains, but it has also presented challenges for human-data interaction. Humans have limited attention spans and may lack the specialized skills to extract valuable insights and craft engaging data stories across multiple modalities~\cite{Li2023c}. Automating the generation of stories from raw data can greatly enhance the efficiency of data analysis and information communication.

Recently, advancements in large language models (LLMs) have showcased robust natural language understanding and reasoning capabilities, proving effective across various tasks like data analysis~\cite{Author2023, Beasley}, document generation~\cite{Lin2023}, and visualization creation~\cite{Dibia2023}. These capabilities open up new avenues to streamline the entire data storytelling workflow by developing systems featuring LLM-powered autonomous agents. In this paradigm, LLMs serve as the cognitive core of these agents, enabling them to perceive environments (Perception), make decisions (Brain), and take responsive actions (Action), thereby assisting humans in automating a wide range of tasks~\cite{Xi2023}. 

Therefore, we aim to explore the potential of LLM-based autonomous agents in facilitating end-to-end storytelling directly from raw data, which is a new problem in the visualization and storytelling community. 
In this paper, we specifically focus on a representative genre of data stories~\cite{Segel2010}, animated data videos, which encompass diverse components and necessitate the coordination of these diverse elements~\cite{wang2022investigating}.
Existing automatic methods for creating data videos either require users to prepare various materials from raw data~\cite{dataplayer, wonderflow, shi_autoclips_2021} or still involve complex and time-consuming manual authoring processes~\cite{DataParticles2023, Thompson2021,ge_cast_2021, Shin2022,amini_authoring_nodate}. 
We envision that autonomous agents can facilitate the automatic transformation of raw data into animated data videos.
However, achieving this goal involves overcoming several challenges:

\begin{figure}[t]
    \centering \includegraphics[width=\linewidth]{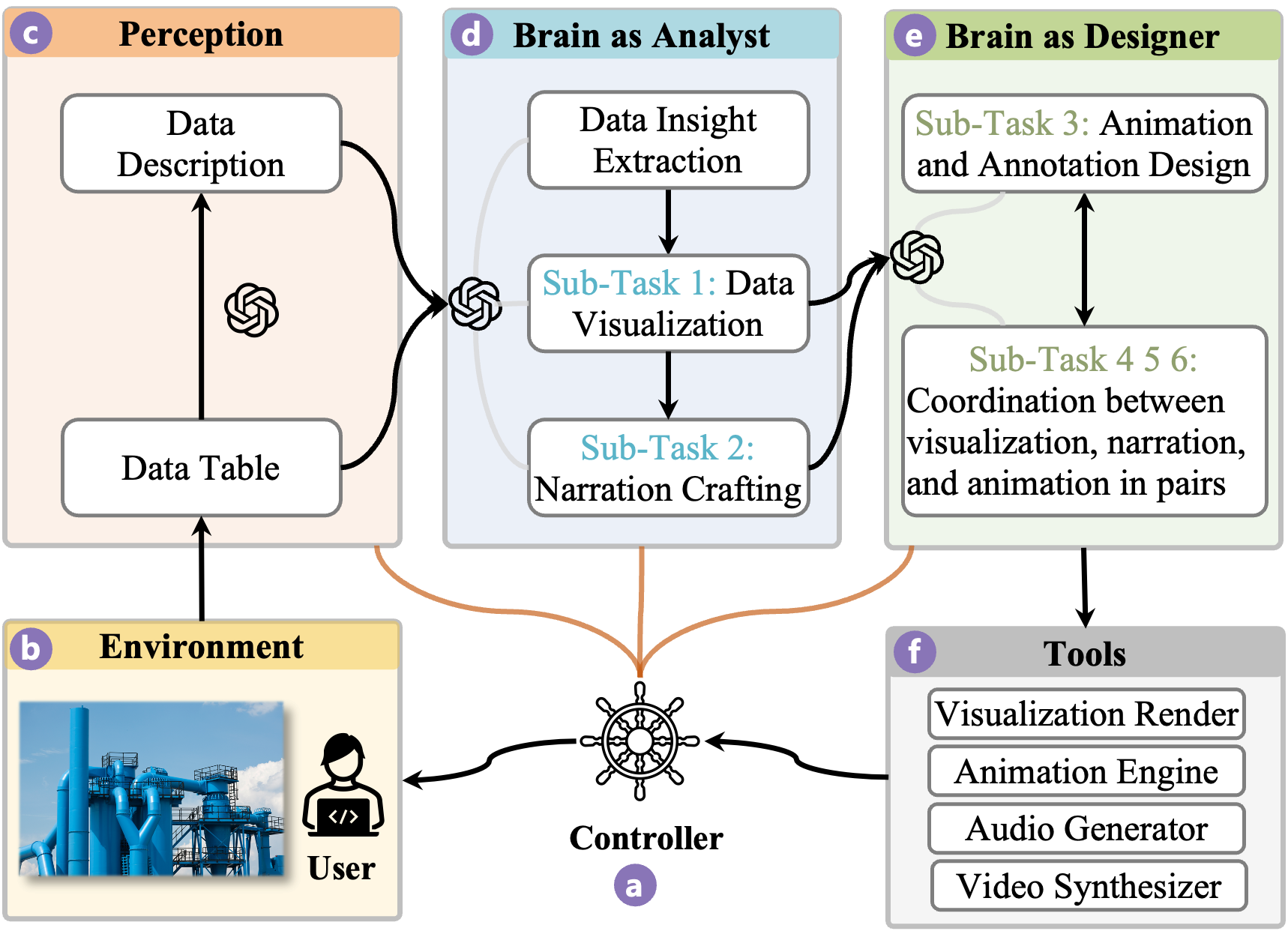}
    \caption{Architecture of \toole. }
    \label{fig:workflow}
\end{figure}

\begin{itemize}
\item \textit{Task Decomposition:}
Data storytelling involves the generation and coordination of diverse elements such as visualizations, text narrations, audio, and animations. The system should accurately interpret raw data, break down the storytelling task into manageable sub-tasks, and assign appropriate roles to agents specialized in handling specific aspects of the task.

\item \textit{Performance Optimization:}
In each sub-task, the agent is required to make informed decisions based on perception inputs and determine the appropriate tools and methods to use. Each sub-task often relies on the outputs of preceding stages, highlighting the interdependence among these sub-tasks. So ensuring optimal performance for each one is crucial.

\item \textit{Workflow Design:}
Storytelling involves numerous interconnected sub-tasks with diverse sequence schemes. Tasks such as data visualization, crafting narration, recording audio, designing animations, and aligning diverse components are typically non-linear and their order may vary, presenting challenges in determining the optimal approach.
The system needs to facilitate a seamless workflow that automates and effectively integrates all these sub-tasks.
\end{itemize}

To better understand these issues, we develop \toole, an LLM-based multi-agent system that automates the entire process of transforming raw data into engaging animated data videos.
The system's architecture is shown in \cref{fig:workflow}.
Specifically, we decompose data video creation into distinct sub-tasks based on data video components and their relationships. We design two agent roles—data analyst and designer—to manage and conduct these sub-tasks, and optimize the performance of each sub-task through prompt design. We also explore effective ways to interconnect these sub-tasks and iteratively refine the workflow.
To demonstrate the effectiveness of \toole, we conduct a case study where \tool generates a data video about real-world stock price data.
Finally, we summarize the lessons learned from our system design for task decomposition, performance optimization, and workflow design, and provide insights to guide future developments in multi-agent systems for the automatic transformation of raw data into data stories.

\begin{figure*}[t!]
    \centering \includegraphics[width=\linewidth]{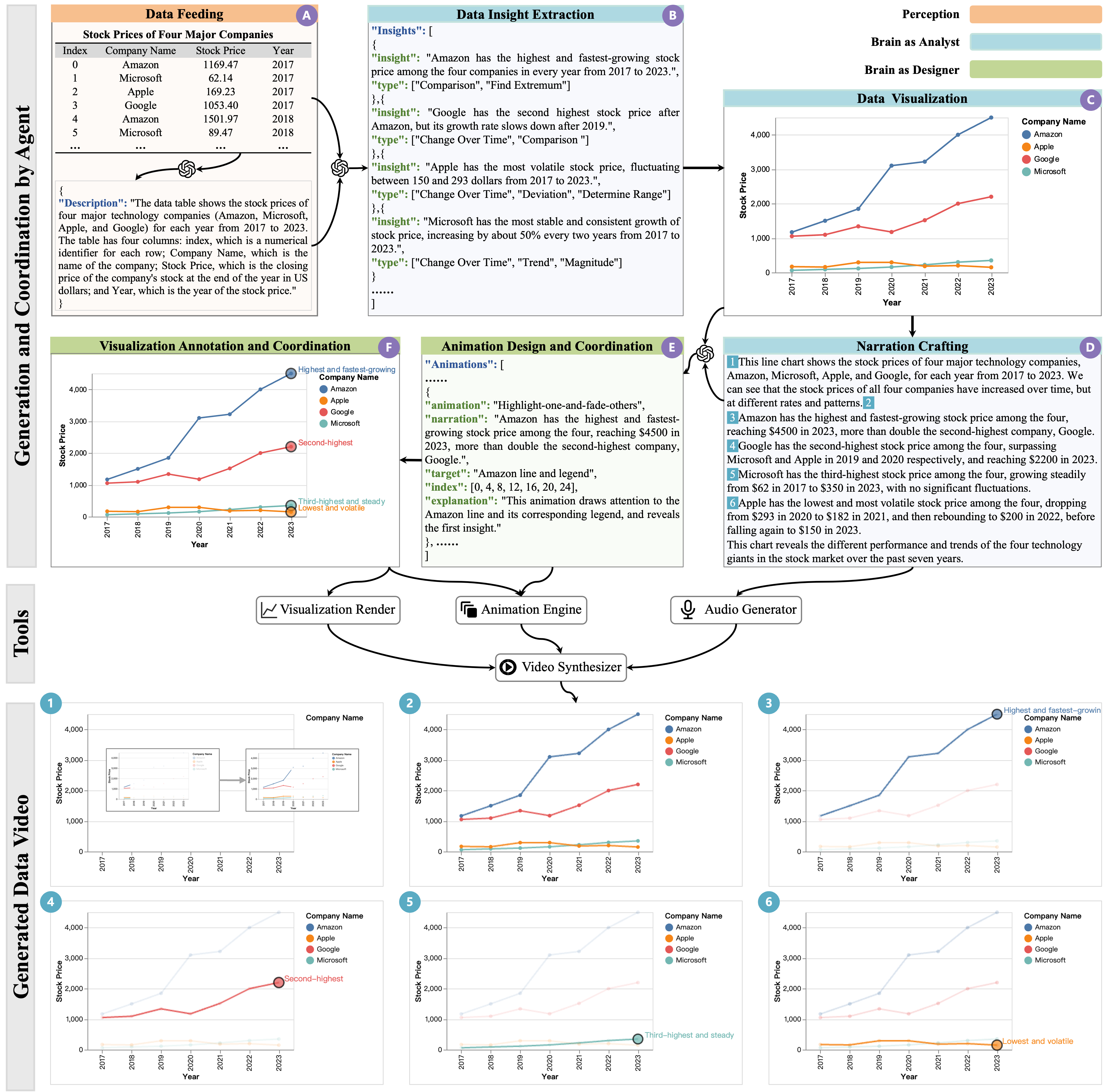}
    \caption{An example walkthrough of \toole. }
    \label{fig:example}
\end{figure*}

\section{\tool}
This section will first give an overview of \toole's architecture and then introduce our design practices.

\subsection{Overview}
\tool is an LLM-based multi-agent system powered by GPT-4~\cite{GPT}. 
We follow existing conceptual framework of LLM-based agent to design three components: perception, brain, and action~\cite{Xi2023}.
The architecture of \tool is illustrated in~\cref{fig:workflow}, with a central controller (a) scheduling all components. User-generated data (b) is directly input into \toole. The perception module (c) preprocesses the data, which is then fed into the first agent acting as a data analyst (d). This agent's tasks include extracting insights, visualizing data, and crafting narration text. The generated visualization and narration text are passed to the next agent, which acts as a designer (e). This agent is responsible for animating and annotating the content, as well as coordinating the data video components. Finally, the controller (a) utilizes the decisions made by the multi-agent system to generate a data video with relevant tools (f).

\subsection{Task Decomposition and Workflow Design}\label{sec:task}
To design multi-agent systems for complex tasks, such as data storytelling, it is crucial to decompose the process into patterned sub-tasks, facilitating model learning and interpretation~\cite{Xi2023}. \textbf{Task decomposition is a balance art between accuracy and efficiency.} Coarse tasks may exceed the model's capabilities, leading to hallucinations or non-computational results, while excessively fine-grained tasks may overwhelm the model with excessive tasks, affecting efficiency and increasing costs.

To decompose the tasks in the context of data video creation, we first break down data videos into basic components and the relationship between these components, following previous research~\cite{wonderflow, dataplayer}.
The basic components of data videos include data visualizations, text narrations, and visual animations.
To make sure these components appear coherently in data videos, three relationships need to be taken care of:
1) Animated \textit{visualization elements} must semantically connect with corresponding \textit{text narration segments};
2) \textit{Animation effects} must be tailored to the \textit{visualization elements} they accompany;
3) \textit{Text narration} should align temporally with the \textit{animations}, serving as the timeline.

With these identified components and relationships, we assign the three component creation tasks and the three relationship management tasks (six sub-tasks in total) to two LLM-powered agents, assuming the roles of a data analyst and a designer.
Additionally, we develop a controller to manage the input and output of each module, parse outputs, and invoke appropriate tools for tasks beyond LLM capabilities. 

\cref{fig:example} illustrates a case study based on real-world stock price data of four IT companies.

\noindent \textbf{Perception.}
The perception module accepts and processes diverse information from external environments, transforming it into understandable representations for LLMs~\cite{Xi2023}.
In \toole, \revise{data tables are inputted directly in the prompt as formatted text.} 
During the perception phase (A), a data preprocessing module utilizes the dataset's title and content within an LLM session to generate natural language descriptions.
As shown in \cref{fig:example}-A, the description involves a high-level overview of the data topic, along with a detailed elaboration of the semantics of each data column, enhancing contextual information for subsequent model processes. 
The generated NL data description and the raw data are then fed into the data analysis brain.

\noindent \textbf{Role as Data Analyst.}
Inspired by insight-based visualization and understanding techniques~\cite{Ying2023, Wang2020i}, when analyzing data, \tool first prompts the LLM to extract top-k interesting data insights (B) from raw data, \revise{based on data analysis task modeling~\cite{shendata, Shen2021}.}
These insights guide the generation of visualizations (sub-task 1) and text narration (sub-task 2). 
Following the Chain-of-Thought strategy~\cite{CoT} and carefully designed LLM prompts, the brain incrementally derives a list of insights (B), declarative Vega-Lite visualizations~\cite{vegalite} (C), and text narrations (D) step by step.

\noindent \textbf{Role as Designer.}
Once the ``data analyst'' has prepared visualizations and corresponding text narrations, the ``designer'' agent focuses on creating dynamic animations (sub-task 3) and synchronizing components (sub-task 4 5 6). 
Animations are categorized into two types: visual effects (\eg fade, grow, fly, zoom, etc.) applied to visualization elements, and annotations that introduce additional visual elements at the right moments. 
Animation effects are determined by distilling choices from an animation library into natural language prompts (E). 
The agent is tasked with selecting the optimal timing for animation applications, identifying the precise visual elements that will be animated, and choosing the appropriate animation effects.
Furthermore, given the complexity of Vega-Lite-specified annotated visualizations, we adopt a hierarchical approach, initially generating base visualizations with the data analyst agent and subsequently enhancing them with annotations (with Vega-Lite specifications) for improved outcomes (F) using the designer agent.
Text narration serves as a timeline, transforming temporal synchronization into semantic links between static narration segments and visual elements~\cite{wonderflow,dataplayer}.

\noindent \textbf{Controller.}
As shown in the middle of \cref{fig:example}, based on the NL outputs of the models, the controller first utilizes text-to-speech services to convert text narrations into audio with precise timestamps, ensuring alignment with the data video timeline. 
Then, the controller invokes a visualization renderer to convert Vega-Lite specifications into SVG files. Each SVG element is automatically associated with blackened data and visualization structure information~\cite{dataplayer, Snyder2023}. Narration segments are further linked to these SVG visual elements based on the text-visual links established by the designer agent, akin to Data Player~\cite{dataplayer}. Then the corresponding animation effects are applied to the SVG elements based on the designer agent's decisions. 
Next, annotated visualizations are parsed to detect SVG annotation elements, integrating ``fade in'' animations at corresponding timestamps. Finally, the controller calls a video synthesizer to merge visualizations, audio narration, and animation sequences into the final data video.


\section{Case Study}
\cref{fig:example} illustrates a real-world case study using stock price data. \tool generates diverse insights (B) and visualizes stock prices over time using line charts (C). The narration followed a structured approach (C): starting with an overview, detailing each company's performance, and concluding with a summary. Based on the narration, points and text annotations were added to each company's line to highlight its notable characteristic (F).
The bottom of \cref{fig:example} displays key animation frames (E) from the video, with each frame numbered to correspond with specific timestamps in the narration (D). This demonstrates when specific animations are triggered. The final data video starts with an entrance animation showcasing all elements, and when discussing each company, the respective line is highlighted individually. Overall, the entire narrative is well-crafted, with smooth articulation and appropriately designed visualizations and animations, resulting in an engaging data video.


\section{Lessons Learned}
This section will discuss the lessons learned throughout the development of \toole, focusing on task decomposition, performance optimization, and workflow design.

\subsection{Task Decomposition}
\noindent \textbf{Balancing Accuracy and Efficiency.}
Task decomposition for data storytelling requires balancing accuracy and efficiency.
When decomposing tasks, 
first, it is essential to identify which tasks the model excels in (\eg natural language generation, reasoning, and text-based decision-making) and distinguish these from tasks that necessitate external tools (\eg visualization rendering, audio generation, and video synthesizing). 
Second, the sub-tasks resulting from decomposition should be well-defined and manageable. These sub-tasks can then be grouped to shape agent roles that align with the inherent characteristics of the tasks. For example, \tool breaks down tasks based on the data video components, and organizes sub-tasks into analysis-focused tasks that generate static content from raw data, and design-focused tasks that require creative input and the derivation of dynamic effects from the static content.
Third, the suitable combination of sub-tasks can enhance both the model's accuracy and efficiency, as demonstrated in \cref{sec:task}, where animation design and temporal synchronization were merged.
Finally, a top-down approach can be adopted to dissect tasks progressively, designing suitable and efficient models for each sub-task.

\noindent \textbf{Data Feeding with Contextual Information.}
Providing the model with ample contextual information has been found to enhance the accuracy and quality of its generation~\cite{Author2023}. 
As the model progresses through sub-tasks step by step, the context is enriched and updated, offering more information for subsequent operations. For example, in the data analyst agent, the model gradually gathers information on data descriptions, insights, visualizations, and narrations.
However, when an agent perceives data from the environment, the data itself is merely numerical with limited context. We have found that semantically enriching the data and supplying the model with contextual insights significantly enhance its effectiveness. For example, \tool uses the LLM to generate an NL description of a data table with a title. 
Future research could integrate innovative techniques for enhancing data comprehension and exploring novel methods for inputting data into LLMs.

\subsection{Performance Optimization}
Effective prompt design is crucial for enhancing the performance and output quality of LLM-based agent systems. The complete prompt of \tool can be found in \cref{sec:prompt}. The key strategies for optimizing prompts within this context are outlined as follows: 

\noindent \textbf{Assignment of Appropriate Tasks for LLMs.}
The foundation of effective prompt design lies in the careful design of tasks for LLMs. It is essential to identify the tasks where LLMs are good at. Furthermore, assigning the model a specific role, such as a data analyst or designer in \toole, can guide the model to produce domain-specific and contextually relevant outputs. 
In addition, supplying the model with precise and comprehensive context can enhance its understanding of tasks, thereby improving the accuracy and relevance of its responses.
This involves crafting well-structured prompts (outlined in \cref{sec:prompt}) and designing complementary modules like data preprocessing in \toole.

\noindent \textbf{Cognitive Processing Time and Task Decomposition.}
Allowing the model adequate cognitive processing time is essential for achieving high-quality outputs and alleviating hallucinations. 
Beyond the task decomposition discussed above, in terms of prompt design, the sub-tasks can be clearly defined with sequential steps, facilitating the application of the Chain-of-Thought strategy. Moreover, the number of tasks within one prompt should be balanced to deduce task difficulty.
In addition, prompting the model to explain its decisions or outline its solution methodology before concluding can promote a more thoughtful and precise response generation process. For instance, in the designer agent, \tool prompts the LLM to its choices regarding the animation and annotation design, which enhances the decision accuracy and simplifies the debugging of LLM applications during development.

\noindent \textbf{Crafting Precise and Unambiguous Instructions.}
The clarity and specificity of instructions are paramount in effective prompt design. Utilizing delimiters (\eg ``` ```, `` '' or $< >$) to segment prompt sections can reduce ambiguity and aid comprehension. Providing fine-grained requirements and employing the correct use of keywords (\eg ``summarize'' vs. ``extract'') ensures that the model adheres closely to the task parameters. Furthermore, requesting structured outputs (\eg JSON and HTML) and offering a range of response options can guide the model toward producing organized and practical outputs. For example, as shown in \cref{fig:example}, \tool allows LLMs to select insight types (B), animation types (E), and annotation types (F) from a set of predefined candidates. Additionally, incorporating conditional logic (\eg if-else statements), employing few-shot or one-shot prompting techniques with curated examples, and referencing URLs for concrete examples can further enhance the model's understanding and task execution accuracy.
More specific examples can be found in \cref{sec:prompt}.

\subsection{Workflow Design}

\noindent \textbf{Shared Representation.}
This paper primarily utilizes the GPT-4 model~\cite{GPT}, with natural language serving as the medium for input and output, setting the stage for this discussion. 
Effective communication between agents and between agents and external tools requires an appropriate shared NL representation. For instance, Vega-Lite is employed in \tool to represent all visualizations and annotations, while insights and animated visual information are encapsulated in a JSON format, incorporating specific feature information (see \cref{fig:example}). Such shared representations must be comprehensible and easily generated by the model and readily interpreted by external tools for mapping to internal operations. Future work could involve designing a global shared representation for specific application scenarios to assist models in better preserving and generating contextual information.

\noindent \textbf{Iterative Development.}
Various sub-tasks within the workflow present diverse sequencing strategies. 
For instance, annotations can be generated simultaneously with visualizations, during animation generation, or using a hierarchical approach as described in \toole. Similarly, in data analysis, one may opt to produce visualizations either before or after narration. 
Designing the optimal sequencing strategy is challenging due to the absence of a quantitative global optimization objective. 
Hence, for applications of LLM-based agents, we adopt an iterative design methodology based on task decomposition and local performance optimization~\cite{Prompt}.
This involves a cycle of ideation, implementation, experimental evaluation, and error analysis. Striving for consistency in the model's outputs also necessitates adherence to established guidelines and meticulous parameter adjustments. 
We note that \tool presented here is the result of our iterative optimizations and may not necessarily represent the optimal configuration. Our intention in developing this prototype tool is to uncover valuable lessons and insights that can guide future research.

\section{Future Work}

\textbf{Global Optimization and Benchmarking.}
The iterative development of multi-agent systems, as mentioned in \cref{sec:task}, suffers from the lack of a global optimization and validation framework for end-to-end data video generation~\cite{Park2023}. Additionally, the community lacks a widely recognized benchmark. The complexity of this challenge is compounded by the inherently subjective nature of data storytelling quality, which is subject to individual interpretation and the multifaceted decision-making involved in various narrative forms. 
Future work could include summarizing relevant rubrics and conducting empirical studies to derive quantitative guidelines. \revise{With these well-defined metrics, an evaluation agent can also be added to enhance existing workflow.}
Additionally, there is a need to develop a universally shared representation for optimization and incorporate domain-specific languages and objectives tailored to diverse scenarios~\cite{NotePlayer,Sallam2022}.

\noindent \textbf{Human-in-the-Loop.}
Data-driven end-to-end generation solutions can result in one-size-fits-all outputs. To address the issues, incorporating human-in-the-loop is an essential approach to compensate for model limitations and generate more personalized results~\cite{Li2023c, li2023ai}. In data storytelling, three paradigms of human-in-the-loop can be further explored: firstly, allowing users to input more information in the perception module while maintaining the current architecture, articulating their goals and requirements in the forms like natural language~\cite{shen2022towards}, example~\cite{Xie2023a,Shen2022b}, and sketch~\cite{Lin2023}; secondly, integrating humans into sub-tasks to achieve local optimization before proceeding to the next stage, such as generating multiple candidates for visualization and annotation after generating data insights; thirdly, users providing conversational feedback based on the output~\cite{shen2022towards,vistalk}, with the agent generating new end-to-end results based on this feedback. Additionally, these methods can also be flexibly combined.

\noindent \textbf{Keeping Up with Cutting-Edge Models.}
This paper primarily uses the GPT-4 model. However, with the rapid evolution of large language models (LLMs), GPT-4 is swiftly being augmented by the emergence of multimodal LLMs~\cite{yin2023survey}. These advanced models offer expanded functionalities for handling multimodal inputs and outputs, significantly impacting task decomposition, performance optimization, and workflow design within our established framework (\cref{fig:workflow}). For instance, initial generation of visualization files could be followed by refinement in a subsequent multimodal module, potentially leading to direct generation of video content.
The enhancement of model capabilities presents numerous opportunities. 
\revise{Future work should not only track the latest models to develop more powerful agents but also leverage diverse models with different capabilities to enrich data storytelling. This includes expanding beyond individual static charts to incorporate visual and musical content~\cite{Tang2022}, supporting more complex insights and multi-view visualizations~\cite{Lin2022}, integrating existing computational design spaces (\eg camera~\cite{Li2023b} and narrative structure~\cite{Yang2022a}), and accommodating more data types (\eg unstructured graphs~\cite{GEGraph}). Achieving these features requires enhancing shared representations and designing corresponding prompts (similar to how \tool integrates animation).}

\noindent \textbf{Inherent Limitations of Large Language Models.}
LLMs are powerful but exhibit several inherent limitations, such as error accumulation, inconsistent results, hallucinations, and high time costs. Most importantly, we need to acknowledge that the content from LLMs is generative but not truthful. To address error accumulation, incorporating a human-in-the-loop approach and providing timely tips can improve accuracy. Consistency in results can be improved by strictly following established guidelines and creating supplementary rules to handle the model's output. Hallucinations, where the model generates plausible but incorrect information, can be improved by implementing some prompt optimization strategies, such as self-repair mechanisms, the Chain-of-Thought (CoT) approach, and code-interpreter functionalities~\cite{Author2023}. Lastly, high-time costs can be managed by breaking down tasks, finding suitable solutions for each (\eg heuristics, basic models, and LLMs).
It's important to recognize that LLMs are not a one-size-fits-all solution; sometimes, basic models or heuristic rules can be highly effective without the need for LLMs.


\section{Conclusion}

The rapid evolution of LLMs presents new opportunities for creating end-to-end multi-agent systems for data storytelling. Through the development of \toole, we have derived valuable insights into task decomposition, local performance optimization through prompt design, and workflow design. In addition, we also shed light on future directions in the development of globally optimized multi-agents, human-in-the-loop systems, integration of cutting-edge multimodal models, and addressing inherent LLM limitations.

\acknowledgments{
The authors wish to thank all reviewers for their valuable feedback.
This work has been partially supported by RGC GRF Grant 16210321. 
}
\bibliographystyle{abbrv-doi}
\balance
\bibliography{main}

\appendix
\onecolumn
\newpage
\section{Full Prompt}\label{sec:prompt}

\begin{lstlisting}[language=Python, caption=The prompt of generating text description for data tables, label=prompt1]

Give a short and consistent description of the following data table and columns:
{{table}}

The title of the data table is: {{title}}

The output JSON format is: 
{
"Description": [A] 
}
where [A] is the generated description.
\end{lstlisting}

\begin{lstlisting}[language=Python, caption=The prompt of the agent
acting as a data analyst, label=prompt2]

You are a data analyst. You have a data table at hand.
{{description}}
The full data table is: 
{{table}}

You need to complete several tasks, please think step by step.

Task 1: Please list the top insights you can gather from the following data table. 
Notes for insight extraction:
- The output JSON format is:
[
    {
	"insight": insight content,
	"type": a list of corresponding insight types
    }
]
- The insight type belongs to the following list: [Change Over Time, Characterize Distribution, Cluster, Comparison, Correlate, Determine Range, Deviation, Find Anomalies, Find Extremum, Magnitude, Part to Whole, Sort, Trend]. One insight can correspond to multiple types.
- The selected insights should be obvious, valuable, and diverse.
- Double-check that the comparison of numerical magnitudes is accurate. Ensure that the insight is right.
- Ignore the "index" column.


Task 2: Please draw a Vega-Lite visualization to tell the insights based on the data table. 
Notes for visualization generation:
- Your title should be less than 10 words.
- If you use the color channel, refer to the following color scheme: https://vega.github.io/vega/docs/schemes/
- Your visualization should have the right height and width.
- If the visualization is a line chart, it should include data points.
- If the focus of the data table presentation is on percentage information of one single column, then use a pie chart to present it. The percentage of each sector should be displayed on the corresponding sector via text annotation like https://vega.github.io/vega-lite/examples/layer_arc_label.html.
- Use a single chart to visualize the data, and the index column should not be visualized.

Task 3: Please write a streamlined narration for the insights.
Notes for narration generation:
- Writing narration in the tone of describing the visualization instead of describing the data table.
- Your logic should be compelling, linking insights into a story, and avoiding enumerating insights.
- Avoid using additional explanations. Avoid speculating on conclusions and redundant explanations beyond the data.
- Content should be streamlined.
- Ignore the "index" column.

The final output JSON format is: 
{
"Insights": [A],
"Visualization": [B],
"Visualization_Type":[C]
"Narration": [D]
}
where [A] is the listed insights in Task 1,  [B] is the visualization specification generated in Task 2, [C] is the visualization type, which is one of the values of [bar, scatter, pie, line], and [D] is the narration text for the insights in Task 3. 

\end{lstlisting}

\begin{lstlisting}[language=Python, caption=The prompt of the agent
acting as a designer, label=prompt3]

You are a data video designer. You have a static visualization and corresponding insightful narration text at hand, as well as the original data table. If necessary, embellish the visualization with corresponding annotations (e.g., arrow, text, circle, etc.) to tell the story more vividly in the narration text. Please think step by step.

The visualization is: {{visualization}}
The insightful narration is: {{narration}}
The data table is: {{table}}

Task 1: You want some animation to appear on the visual elements during the audio narration of the video to aid in telling the data story.  Consider the narration as a timeline for the video. Insert animations inside the narration text where you feel they are needed. The corresponding animation will be triggered when the video reaches the corresponding narration segment and ends at the end of the narration segment.
Notes for animation generation:
- The output JSON format is:
[
  {
    "animation": animation type,
    "narration": narration segment,
    "target": the visual elements that the animation applies to,
    "index": a list of related data table rows index. If empty, output [],
    "explanation": the explanation of why the animation needs to be added
  }
]
- One sentence in the narration can correspond to multiple animations.
- Output only the part marked with animation.
- Narration text cannot be modified.
- There are three types of animations: entrance, emphasis, and exit.
- Entrance animations include [Axes-fade-in, Bar-grow-in, Line-wipe-in, Pie-wheel-in, Pie-wheel-in-and-legend-fly-in, Scatter-fade-in, Bar-grow-and-legend-fade-in, Line-wipe-and-legend-fade-in, Fade-in, Float-in, Fly-in, Zoom-in].
- Emphasis animations include [Bar-bounce, Zoom-in-then-zoom-out, Shine-in-a-short-duration, Highlight-one-and-fade-others].
- Exit animations include [Fade-out].
- [Axes fade in] animations can only be used at the beginning (first sentence) of a whole narration.
- Visual elements that have an entrance animation effect applied will not appear on the canvas until the animation is triggered. Elements that have an exit animation effect applied will disappear from the canvas after the animation. Elements that do not have any animation effect applied will appear on the canvas by default. 
- Visual elements can only be emphasized or disappear after they appear on the canvas, and elements cannot be emphasized after they disappear.

Task 2: Visualization embellishment generation
Notes for generation:
- Add annotation only when you think you need to, or export the original visualization if you do not feel the need to add it.
- Tag the narration text in the format of:
[
  {
    "type": a list of annotation types, which is one or a set from [mark label, circle, text, rule, trend line, arrow],
    "description": annotation description and explanation,
    "index": a list of related data table rows index, If empty, output [],
    "nar": narration segment
  }
]
- The tagged annotations must correspond to the annotations in the visualization.
- Only output the narration segments marked with annotation. If the value of key "type" is [], do not output the item. 
- If the annotated vega-lite specification has a key "layer", then all "mark" and "encoding" keys should be inside the list value of the key "layer".
- The annotations should not be complex.
- The annotation must correspond to the narration segment. The annotation will appear when the video reaches the corresponding narration segment.
- The text annotation should be short (e.g., fewer than 6 words).
- Please output the complete bug-free vega-lite specification.

The final output JSON format is: 
{
"Annotated_Visualization": [A],
"Annotated_Narration_for_Animation": [B],
"Annotated_Narration_for_Annotation": [C]
}
where [A] is the generated annotated Vega-Lite specification, [B] is the tagged narration text for animation, and [C] is the tagged narration text for annotation.

\end{lstlisting}

\end{document}